\documentclass[11pt]{article}
\usepackage{graphicx}

\begin{document}

\title{Polaron Properties of an Impurity in Bose-Einstein-Condensation}

\author{Beibing Huang and Shaolong Wan\thanks{Corresponding author.
Electronic address: slwan@ustc.edu.cn} \\
Institute for Theoretical Physics and Department of Modern Physics \\
University of Science and Technology of China, Hefei, 230026, {\bf
P. R. China}}

\maketitle
\begin{center}
\begin{minipage}{120mm}
\vskip 0.8in
\begin{center}{\bf Abstract} \end{center}

{In this paper we study an impurity in Bose-Einstein-Condensate
system at $T=0K$ and suppose the contact forms for boson-boson and
boson-impurity interactions. Using Bogoliubov theory and a further
approximation corresponding to only think over the forward
scattering of impurity by bosons, we derive a reduced Hamiltonian
whose form is the same as the Fr\"{o}hlich Hamiltonian for large
polaron. By using Lee-Low-Pines (LLP) theory for large polaron, we
obtain the effective mass of impurity, the phonon number carried by
impurity and the energy related to the existence of impurity. In
addition, we also discuss the valid range for forward-scattering
approximation.}

\end{minipage}
\end{center}

\vskip 1cm

\textbf{PACS} number(s): 03.75.Kk, 03.75.Ss, 74.20.Mn
\\
\\

\section{Introduction}

Since the impressive experimental achievement of
Bose-Einstein-Condensation (BEC) \cite{M.H., Davis, Bradley} quantum
degenerate atomic gases have become one of the hottest domains.
Temperature in such systems can be thought to approach zero so that
the thermal fluctuations and classical phase transitions associated
with thermal fluctuations are fully suppressed. But quantum
fluctuations due to Heisenberg uncertainty principle still exist and
may be strong enough to lead to a quantum phase transition. So these
systems are good playground to realization of quantum phase
transitions \cite{M.Greiner, fisher, jaksch}. Except probing for
fundamental physics in the realm of dilute atomic quantum gas, BEC
system is also an auxiliary but effective tool to sympathetically
cool some systems, which can not be cooled by traditional cooling
technique, to degenerate temperature \cite{myatt, schreck,
truscott}. So a quantum degenerate mixing system could be obtained.
Extending such mixing system to a extreme case that there is only
one atom belonging to sympathetically cooled system, we will expect
a BEC system with an impurity. Fortunately, such system has been
able to create in experiment by many ways \cite{truscott, a.p.,
hadzibabic, roati}. An impurity in BEC is very similar to polaron,
an electron in ionic crystal. When the electron moves, it polarizes
the lattice around it and inversely experiences an effective
potential from the lattice. And in the BEC system, the existence of
an impurity would modify the distribution of atoms in BEC and also
produce an effective potential to act on itself \cite{f.m.}. In
\cite{f.m., ryan, k.s.}, their authors made good use of the product
wave function similar to Landau-Pekar description to polaron
\cite{landau, pekar} to deal with the self-localization of impurity.
The product wave function, the essence of which is the separation of
freedoms, is a good approximation for polaron owing to large mass
ratio of ion to the electron to make the Born-Oppenheimer
approximation be valid. But the validity of product wave function is
questionable in the BEC system with an impurity, owing to similar
masses between the bosons and the impurity, and some deviations from
it can be expected \cite{k.s.}.

In this article, we start from the bosonic system Hamiltonian with
an impurity, and by using Bogoliubov theory, derive an equivalent
Hamiltonian which is the same as the Fr\"{o}hlich Hamiltonian
\cite{fr1, fr2} for large polaron in section 2. In this equivalent
Hamiltonian, Bogoliubov phonon plays a role of the phonon which
comes from the lattice oscillation. Then we use intermediate
coupling theory \cite{Lee} of Lee-Low-Pines (LLP) for polaron to
calculate the effective mass of impurity, the phonon number carried
by impurity and the impurity energy in section 3. Finally, some
discussions about validity of our approximate decision on condensate
fraction and conclusions are given in section 4.

\section{Equivalent Hamiltonian for the Bosonic System with an Impurity}

We consider a single impurity immersed in a homogeneous BEC in three
dimension space. Assuming that the boson-boson and boson-impurity
interactions can be described by contact interactions, the many-body
Hamiltonian \cite{ryan} reads
\begin{eqnarray}
H &=& - \sum_{i} \frac{\hbar^{2} \bigtriangledown_{i}^{2}}{2m_{B}}
+
\frac{g}{2} \sum_{i,j} \delta(\vec{r}_{i} - \vec{r}_{j}) \nonumber \\
&&+ \lambda \sum_{i} \delta(\vec{r} - \vec{r}_{i}) - \frac{\hbar^2
\bigtriangledown^{2}}{2m_{I}} \label{1.0}
\end{eqnarray}
where $m_B$, $m_I$ are the masses of boson and impurity
respectively. $g=4\pi\hbar^2a_{BB}/m_B$,
$\lambda=2\pi\hbar^2a_{IB}/m_{red}$ are the interaction strength of
contact potentials for boson-boson and boson-impurity interaction
with the reduced mass $m_{red}=m_{I}m_{B}/(m_{I}+m_{B})$.
$\vec{r}_i$, $\vec{r}$ represent the coordinates of bosons and
impurity, respectively. In the second quantization representation,
we can expand the bosonic field operator on a set of plane wave
basis and change Hamiltonian (\ref{1.0}) into
\begin{eqnarray}
H &=& \sum_{k}(\varepsilon_{k} - \mu) a_{k}^{\dag} a_{k} +
\frac{g}{2} \sum_{k, p, q} a_{k}^{\dag} a_{p}^{\dag} a_{p - q} a_{k + q} \nonumber \\
&&+ \lambda \sum_{k, q} a_{k}^{\dag} a_{q} \exp \left[- i (\vec{k}
- \vec{q}) \cdot \vec{r} \right] + \frac{P^2}{2 m_{I}} \label{1.1}
\end{eqnarray}
where $\varepsilon_{k}=\hbar^{2}k^{2}/(2m_{B})$ and $\vec{P}$ is the
momentum of impurity. The chemical potential $\mu$ is introduced to
keep the average bosonic number to be a constant. $a_{k},
a_{k}^{\dag}$ are the bosonic annihilation and creation operators
conforming to the canonical commutation relation $[a_{k},
a_{q}^{\dag}]=\delta_{kq}$. Owing to only one impurity, the
statistical property of impurity is unimportant and the same results
are obtained for the fermionic and bosonic impurity. For the
convenience, we suppose our system to be unit volume.

At $T=0K$, the bosons macroscopically occupy the lowest energy
state with wave vector $\vec{k}=0$, which makes it possible to
neglect quantum property of $a_{0}, a_{0}^{\dag}$ and consider
them to be classical number \cite{van}. Following the general
procedure, we decompose $a_{k}$ as follows
\begin{eqnarray}
a_{k} \rightarrow \sqrt{n_0} \delta_{k0} + a_{k}  \label{1.2}
\end{eqnarray}
$n_{0}$ is the number density of atom on the lowest energy state
with $n_{0}=a_{0}^{\dag}a_{0}$. Substituting (\ref{1.2}) into
(\ref{1.1}) and only keeping terms until the second order, the
Hamiltonian reads
\begin{eqnarray}
H &=& \frac{P^2}{2 m_{I}} - \mu n_{0} + \frac{g}{2} n_{0}^{2} +
\lambda n_{0} + \sqrt{n_{0}} (- \mu + g n_{0} + \lambda)(a_{0} +
a_{0}^{\dag})  \nonumber \\
&&+ \lambda \sqrt{n_{0}} \sum_{k}^{'} \left[ a_{k} \exp(i \vec{k}
\cdot \vec{r}) + a_{k}^{\dag} \exp(- i \vec{k} \cdot \vec{r})
\right]+ \lambda \sum_{k, q} a_{k}^{\dag} a_{q} \exp[- i (\vec{k} -
\vec{q}) \cdot \vec{r}]\nonumber \\
&& + \sum_{k} (\varepsilon_{k} - \mu) a_{k}^{\dag} a_{k} + \frac{g
n_{0}}{2} \sum_{k} (a_{k} a_{-k} + 4 a_{k}^{\dag} a_{k} +
a_{k}^{\dag} a_{-k}^{\dag}) \label{1.3}
\end{eqnarray}
where the sum with character $\sum_{k}^{'}$ represents exclusion for
$\vec{k} = 0$ state. Note that both interactions of boson-boson and
boson-impurity make the BEC deplete. We take into consideration
contributions of such two kinds of interaction to the condensate
depletion at the same time by asking for the coefficient of $a_{0},
a_{0}^{\dag}$ to vanish \cite{van} and find
\begin{eqnarray}
\mu = g n_{0} + \lambda  \label{1.4}
\end{eqnarray}
When BEC happens, the gauge symmetry of the system is broken and a
long-wavelength gapless mode exists which is imposed by Goldstone
theorem \cite{goldstone}. In order to satisfy above condition, we
find that if we make below approximation for boson-impurity
scattering term
\begin{eqnarray}
\lambda \sum_{k, q} a_{k}^{\dag} a_{q} \exp{[- i (\vec{k} - \vec{q})
\cdot \vec{r}]} \approx \lambda \sum_{k} a_{k}^{\dag} a_{k}
\label{1.5}
\end{eqnarray}
the gapless property of low energy excitations is met as we could
see below. Physically this approximation implies that when the
impurity is scattered by bosons, the momentum of bosons and impurity
will not alter. In other words, impurity only experiences forward
scattering. So the Hamiltonian is further reduced to
\begin{eqnarray}
H &=& \frac{P^2}{2 m_{I}} - \frac{g}{2} n_{0}^{2} - \frac{1}{2}
\sum_{k} (\varepsilon_{k} + g n_{0}) \nonumber \\
&& + \lambda \sqrt{n_{0}} \sum_{k}^{'} \left[ a_{k} \exp{(i
\vec{k}
\cdot \vec{r})} + a_{k}^{\dag} \exp{(-i\vec{k} \cdot \vec{r})} \right] \nonumber \\
&& + \frac{1}{2} \sum_{k} (\varepsilon_{k} + g n_{0})(a_{k}^{\dag}
a_{k} + a_{-k} a_{-k}^{\dag}) \nonumber \\
&& + g n_{0}(a_{k} a_{-k} + a_{k}^{\dag} a_{-k}^{\dag})
\label{1.6}
\end{eqnarray}
Making the Bogoliubov transformation
\begin{eqnarray}
a_{k} &=& u_{k} b_{k} - v_{k} b_{-k}^{\dag} \nonumber \\
a_{-k}^{\dag} &=& - v_{k} b_{k} + u_{k} b_{-k}^{\dag} \label{1.7}
\end{eqnarray}
and imposing the conditions
\begin{eqnarray}
&& u_{k}^{2} - v_{k}^{2} = 1 \nonumber \\
&& (u_{k}^{2} + v_{k}^{2}) g n_{0} - 2 u_{k} v_{k} (g n_{0}+
\varepsilon_{k}) = 0 \nonumber \\
&& \xi_{k} = (u_{k}^{2} + v_{k}^{2})(\varepsilon_{k} + g n_{0}) -
2 u_{k} v_{k} g n_{0} \label{1.8}
\end{eqnarray}
We obtain
\begin{eqnarray}
H &=& E_{B} + \frac{P^2}{2 m_{I}} + \sum_{k} \xi_{k} b_{k}^{\dag}
b_{k} \nonumber \\
&+& \sum_{k}^{'} \left[ V_{k} b_{k} \exp{(i \vec{k} \cdot \vec{r})}
+ V_{k}^{\ast} b_{k}^{\dag} \exp{(- i \vec{k} \cdot \vec{r})}
\right] \label{1.9}
\end{eqnarray}
where $V_{k} = \lambda \sqrt{n_{0}}(u_{k} - v_{k})$ and $E_{B} = -
\frac{g}{2} n_{0}^{2} + \frac{1}{2} \sum_{k}(\xi_{k} -
\varepsilon_{k} - g n_{0})$. $\xi_{k} = \sqrt{\varepsilon_{k}^{2} +
2 g n_{0} \varepsilon_{k}}$ with $v_{k}^{2} = u_{k}^{2} - 1 =
\frac{1}{2}(\frac{\varepsilon_{k} + g n_{0}}{\xi_{k}} - 1), u_{k}
v_{k} = \frac{g n_{0}}{2 \xi_{k}}$. It is very clear that low energy
excitation $\xi_{k}$ is gapless and completely same as that in the
situation without the impurity. $n_{0}$ is decided by atomic number
conservation
\begin{eqnarray}
n &=& n_{0} + \sum_{k}<a_{k}^{\dag} a_{k}> \nonumber \\
&=&n_{0}+\sum_{k}v_{k}^{2} + \sum_{k} (u_{k}^{2} + v_{k}^{2})
<b_{k}^{\dag} b_{k}> - \sum_{k} u_{k} v_{k} <b_{k}^{\dag}
b_{-k}^{\dag} + b_{k} b_{-k}> \label{1.10}
\end{eqnarray}
Where $<\cdots>$ represents the ensemble average. In the pure
bosonic system, Bogoliubov transformation makes the Hamiltonian of
bosonic system diagonal about phonon operators, so the last two
terms on the right hand side disappear and $n_{0}$ is decided by
equation
\begin{eqnarray}
n = n_{0} + \sum_{k} v_{k}^{2} \label{1.100}
\end{eqnarray}
While for the bosonic system with an impurity the last two terms are
nonzero, leading to that the particle number equation and bosonic
condensate fraction are modified in contrast to pure bosonic system.
In fact, the last two terms in (\ref{1.10}) illustrate the
contribution of impurity to condensate depletion. But as we will see
below, the last two terms are much smaller than other terms in
(\ref{1.10}) and are negligible due to small ratio of impurity
number to boson number. Thus, we determine condensate fraction
according to (\ref{1.100}), in other words, we do not consider the
contribution of boson-impurity interaction to condensate depletion.
When $n_{0}$ is determined, the Hamiltonian (\ref{1.9}) is
completely decided.

Owing to nonexistence of zero-momentum phonon, exclusion for
$\vec{k} = 0$ state in (\ref{1.9}) is not important and we can
neglect this exclusion. In addition to a constant term $E_B$, the
form of Hamiltonian (\ref{1.9}) is the same as that of Fr\"{o}hlich
large polaron theory. Bogoliubov phonon is akin to phonon coming
from lattice oscillations. When an electron moves in ionic crystal,
it will be influenced more strongly by optical phonon, which
represents relative movement of positive and negative ions and is
accompanied by polarized electric field, than acoustic one which
represents the movement of mass center and can not produce polarized
electric field. So in usual Fr\"{o}hlich large polaron theory, the
phonon is considered to be optical mode one and its frequency is
chosen to be independent of the wavevector and be a nonzero constant
due to finite energy gap for optical phonon. But in BEC system with
an impurity phonon is gapless, moreover we must consider the
dependence of dispersion relation on wavevector. As a summarization
of this section, we get an effective Hamiltonian similar to
Fr\"{o}hlich large polaron by making Bogoliubov and
forward-scattering approximations.

\section{The Effective Mass, Phonon Number and Impurity Energy}

Below, we follow the LLP approach \cite{Lee} to calculate some
properties of such system. The essence of the LLP theory consists in
combining the canonical transformation with the variational
principle to get approximate ground state of the system. Supposing
$|\Phi>$ is the ground state $H |\Phi> = E |\Phi>$, we make the
canonical transformation $|\Phi>=U_{1}|\Psi>$ with
$U_{1}=\exp{[-i\sum_k b_{k}^{\dag}b_{k} \vec{k} \cdot \vec{r}]}$ and
the Hamiltonian is transformed into $H^{'}=U_{1}^{-1}HU_{1}$
\begin{eqnarray}
H^{'} &=& E_{B} + \frac{(\vec{P} - \sum_{k} \hbar \vec{k}
b_{k}^{\dag} b_{k})^{2}} {2 m_{I}} \nonumber \\
&& + \sum_{k} \xi_{k} b_{k}^{\dag} b_{k} + \sum_{k} (V_{k} b_{k} +
V_{k}^{\ast} b_{k}^{\dag}) \label{1.11}
\end{eqnarray}
In (\ref{1.11}), the coordinate of impurity is expunged and its
momentum becomes a good quantum number. In fact, the momentum
$\vec{P}$ in (\ref{1.11}) represents the whole momentum of the
system. Introducing the variational function $f(k)$ and making
another canonical transformation $|\Psi> = U_{2} |0>$ with $U_{2} =
\exp{[\sum_{k} (b_{k}^{\dag} f(k) - b_{k} f^{\ast}(k))]}$ and $|0>$
representing the vacuum of the phonon, the Hamiltonian (\ref{1.11})
can be rewritten as $\widehat{H} = U_{2}^{-1} H^{'} U_{2}$
\begin{eqnarray}
\widehat{H} &=& E_{B} + \frac{(\vec{P} - \sum_{k} \hbar \vec{k}
b_{k}^{\dag} b_{k})^{2}} {2 m_{I}} + \sum_{k} \left[ V_{k} f(k) +
V_{k}^{\ast} f^{\ast}(k) \right] + \frac{\hbar^{2}} {2 m_{I}}
\sum_{k, q} \vec{k} \cdot \vec{q} |f(k)|^{2} |f(q)|^{2} \nonumber \\
&+& \sum_{k} |f(k)|^{2} \left[\epsilon_{k} - \frac{\hbar} {m_{I}}
\vec{P} \cdot \vec{k} + \xi_{k} \right] + \sum_{k} \xi_{k}
b_{k}^{\dag} b_{k} + \sum_{k }b_{k}^{\dag} b_{k} \sum_{q}
\frac{\hbar^{2}} {m_{I}} \vec{k} \cdot \vec{q} |f(q)|^{2} \nonumber \\
&+& \sum_{k} \left\{ b_{k}^{\dag} \left[ V_{k}^{\star} + f(k)
\left(\xi_{k} + \epsilon_{k} - \frac{\hbar}{m_{I}} \vec{P} \cdot
\vec{k} + \frac{\hbar^{2}}{m_{I}} \sum_{q} \vec{k}
\cdot \vec{q} |f(q)|^{2} \right) \right] + h.c. \right\} \nonumber\\
&+& \frac{\hbar^{2}}{2 m_{I}} \sum_{k, q} \vec{k} \cdot \vec{q}
\left[ b_{k}^{\dag} b_{q}^{\dag} f(k) f(q) + 2 b_{k}^{\dag} b_{q}
f(k) f^{\ast}(q) + b_{k} b_{q} f^{\ast}(k) f^{\ast}(q)
\right]\nonumber \\
& + & \frac{\hbar^{2}}{m_{I}} \sum_{k, q}\vec{k} \cdot \vec{q}
\left[ f^{\ast}(q) b_{k}^{\dag} b_{k} b_{q} + f(q) b_{q}^{\dag}
b_{k}^{\dag} b_{k} \right]  \label{1.12}
\end{eqnarray}
where $\epsilon_{k} = \hbar^{2} k^{2}/(2 m_{I})$. The ground state
energy of the system $E_{g}(P) = <0|\widehat{H}|0> = E_{B} + E(P)$,
where
\begin{eqnarray}
E(P)&=&\frac{P^{2}}{2 m_{I}} + \sum_{k} \left[V_{k} f(k) +
V_{k}^{\ast} f^{\ast}(k) \right] + \sum_{k} |f(k)|^{2} \left[
\epsilon_{k} - \frac{\hbar}{m_{I}} \vec{P} \cdot \vec{k} + \xi_{k}
\right] \nonumber \\
&&+ \frac{\hbar^{2}}{2 m_{I}} \left[ \sum_{k} \vec{k} |f(k)|^{2}
\right]^{2} \label{1.13}
\end{eqnarray}
and $f(k)$ is decided by $\frac{\delta E(P)}{\delta f(k)}=
\frac{\delta E(P)}{\delta f^{\ast}(k)} = 0$
\begin{eqnarray}
V_{k} + f^{\ast}(k) \left[\xi_{k} + \epsilon_{k} -
\frac{\hbar}{m_{I}} \vec{P} \cdot \vec{k} + \frac{\hbar^{2}}{m_{I}}
\sum_{q}|f(q)|^{2}\vec{k}\cdot\vec{q} \right] = 0 \nonumber \\
\label{1.14}
\end{eqnarray}

According to the LLP approach, letting $\sum_{k} \hbar \vec{k}
\mid f(k) \mid^{2} = \eta \vec{P}$, we have
\begin{eqnarray}
f^{\ast}(k) = - \frac{V_{k}}{\xi_{k} + \epsilon_{k} -
\frac{\hbar}{m_{I}}(1 - \eta) \vec{P} \cdot \vec{k}} \label{1.15}
\end{eqnarray}
After introducing the parameter $\eta$, the behavior of the system
completely depends on the value of $\eta$. We further have
\begin{eqnarray}
\eta \vec{P} = \sum_{k} \frac{\hbar \vec{k} |V_{k}|^{2}}
{\left[\xi_{k} + \epsilon_{k} - \frac{\hbar}{m_{I}}(1 - \eta)
\vec{P} \cdot \vec{k} \right]^{2}} \label{1.16}
\end{eqnarray}
The parameter $\eta$ is self-consistently decided in (\ref{1.16}),
which is the main equation in LLP theory. Substituting (\ref{1.15})
into the expression of $E(P)$
\begin{eqnarray}
E(P) &=& \frac{P^{2}}{2 m_{I}}(1 - \eta^{2}) - \sum_{k}
\frac{|V_{k}|^{2}} {\xi_{k} + \epsilon_{k} - \frac{\hbar}{m_{I}}(1 -
\eta) \vec{P} \cdot \vec{k}}  \label{1.17}
\end{eqnarray}
Below we make a small quantity expansion about $P$ to (\ref{1.16})
till the second order to get
\begin{eqnarray}
\eta \vec{P} = \sum_{k} \frac{\hbar\vec{k}|V_{k}|^{2}}
{(\xi_{k}+\epsilon_{k})^{3}}\frac{2\hbar}{m_{I}}(1-\eta) \vec{P}
\cdot\vec{k} \label{1.18}
\end{eqnarray}
Choosing the orientation of $\vec{P}$ to be along the z axis, we
have
\begin{eqnarray}
\frac{\eta}{1-\eta}=\sum_{k} \frac{|V_{k}|^{2}}
{(\xi_{k}+\epsilon_{k})^{3}}\frac{2\hbar^{2}}{m_{I}}k^{2}
\cos^{2}\theta \label{1.19}
\end{eqnarray}
In order to calculating the effective mass of impurity, we also
expand the $E(P)$ to second order of $P$ and find
\begin{eqnarray}
E(P) &=& - \sum_{k} \frac{\mid V_{k} \mid^{2}} {\xi_{k} +
\epsilon_{k}} + \frac{P^{2}}{2 m_{I}} (1 - \eta^{2}) \\
&& - \frac{P^{2}}{2 m_{I}} (1 - \eta)^{2} \sum_{k} \frac{\mid
V_{k} \mid^{2}} {(\xi_{k} + \epsilon_{k})^{3}} \frac{2
\hbar^{2}}{m_{I}} k^{2} \cos^{2} \theta \nonumber \label{1.20}
\end{eqnarray}
By means of (\ref{1.19}), the effective mass of the impurity
$m_{I}^{\ast}$
\begin{eqnarray}
m_{I}^{\ast} = \frac{m_{I}}{1 - \eta} \label{1.21}
\end{eqnarray}
Introducing another parameter $W=\eta/(1-\eta)$ which, in fact, can
reflect the effective strength of coupling in contrast to the
situation for polaron, the effective mass can be expressed into
$m_{I}^{\ast}/m_{I}=1+W$. Integrating the angular variable, $W$ is
reduced as
\begin{eqnarray}
W = \frac{8\rho}{3\sqrt{3}\pi^{3/2}} \frac{m_{B}}{m_{I}} \left(1 +
\frac{m_{B}}{m_{I}} \right)^{2} \frac{(k_B a_{IB})^{2}}{
(k_Ba_{BB})^{1/2}} \Pi^{(0)} \label{1.22}
\end{eqnarray}
where
\begin{eqnarray}
\Pi^{(0)} = \int_{0}^{\infty}dy \frac{y^{2}} {(y^{2} +
2\rho)^{1/2}[(y^{2} + 2 \rho)^{1/2} + (m_{B}/m_{I}) y]^{3}}
\nonumber \\ \label{1.23}
\end{eqnarray}
Where in order to scale the expression we define a parameter $k_{B}$
like the Fermi wavevector $n=k_{B}^3/(6\pi^2)$. In addition, we also
introduce another notation $\rho=n_{0}/n$ for condensate fraction.
From the expression of $W$ and under the approximation (\ref{1.100})
which signifies the condensate fraction $\rho$ is independent of
$a_{IB}$, we easily find that $W$ is proportional to the square of
the $a_{IB}$ for definite the $a_{BB}$ and $m_{B}/m_{I}$. Fig.1(a)
apparently respects this kind of behavior. The dependence of $W$ on
$m_{B}/m_{I}$ for definite $a_{BB}$ and $a_{IB}$ appears to be
indirect due to the dependence of $\Pi^{(0)}$ on $m_{B}/m_{I}$.
Fig.1(b) show the corresponding change of the effective coupling
strength as function of $m_{B}/m_{I}$.

\begin{figure}[tbp]
\centering
\includegraphics [width=7.5cm, height=6.0cm]{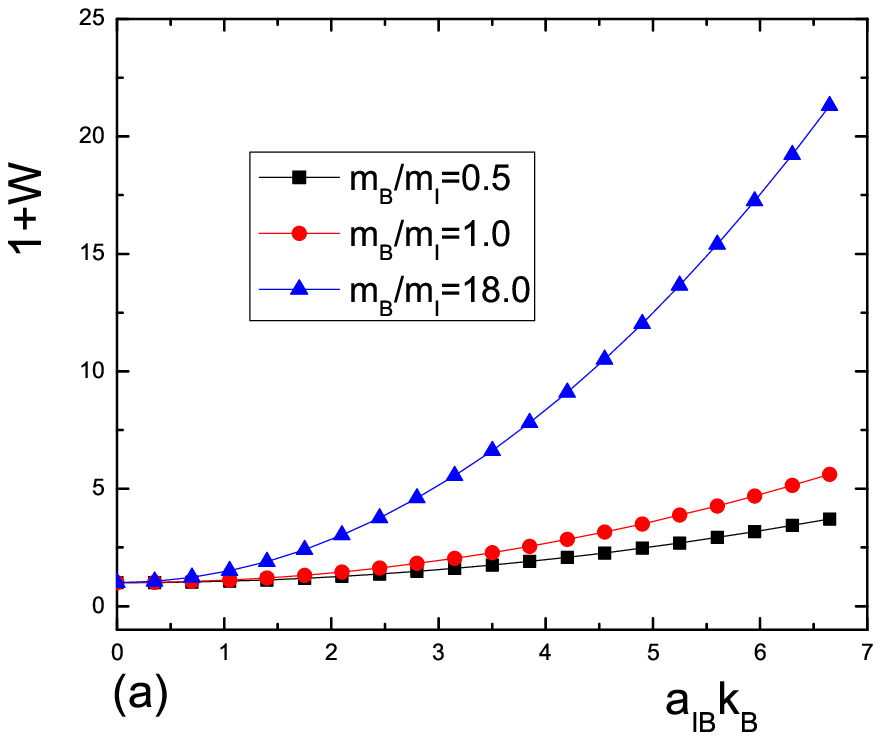}
\includegraphics [width=7.5cm, height=6.0cm]{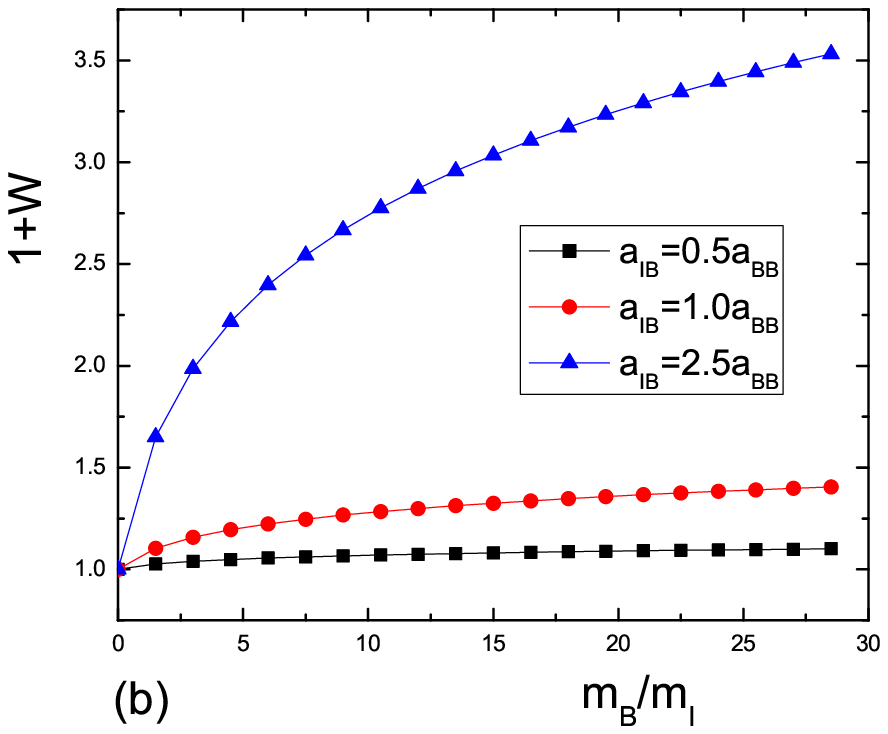}
\caption{The behavior of the effective mass $m_{I}^{\ast}/m_{I}=1+W$
as a function of $a_{IB}$ for different mass ratios $m_{B}/m_{I}$ of
atom to impurity (a); as a function of mass ratio $m_{B}/m_{I}$ for
different boson-impurity scattering length $a_{IB}$ (b). We have
chosen $a_{BB}k_{B}=0.87$.} \label{fig.1}
\end{figure}

At zero temperature, the number of phonon is zero for purely BEC
while there exists some phonon excitations in BEC with an impurity.
Below we calculate the phonon number carried by the impurity
$n_{ph}=\sum_{k}<\Phi|b_{k}^{\dag}b_{k}|\Phi> = \sum_{k}|f(k)|^{2}$
\begin{eqnarray}
n_{ph}
=\frac{2\rho}{\sqrt{3}\pi^{3/2}}\frac{(k_Ba_{IB})^2}{(k_Ba_{BB})^{1/2}}
\left(1 + \frac{m_{B}}{m_{I}} \right)^{2}\Pi^{(1)} \label{1.24}
\end{eqnarray}
where
\begin{eqnarray}
\Pi^{(1)}=\int_{0}^{\infty} d y \frac{y}{(y^{2} + 2 \rho)^{1/2}}
\frac{1}{[(y^{2} + 2\rho)^{1/2} + (m_{B}/m_{I}) y]^{2} - 4
(m_{B}/m_{I})^2 (1 - \eta)^{2} \widetilde{P}^2} \label{1.25}
\end{eqnarray}
and $\widetilde{P} = P/(2m_B n g)^{1/2}$. When an impurity moves in
BEC, the state of this impurity is sensitively dependent on its
velocity according to superfluid Landau theory \cite{lll}. Landau
used simple kinematic arguments to derive an expression for the
critical velocity $V_L=min[\zeta(k)/k]$ where $\zeta(k)$ is the
energy of an elementary excitation with momentum $k$. If the
velocity of impurity is larger than $V_L$, the impurity will
dissipate energy by colliding with BEC, but if not, the impurity
will move without dissipation. Such behavior has been predicted
theoretically \cite{timmerse} and observed experimentally
\cite{a.p.}$\colon$ as the impurity velocity is diminished below
Landau critical velocity, there exists a dramatic reduction in the
probability of collisions. Recalling that in the process of deriving
effective Hamiltonian (\ref{1.9}) we have made forward-scattering
approximation. Only forward scattering corresponds to no scattering
and vanishing collision probability. Thus we think that the
reduction of collision probability for low impurity speed smaller
than Landau critical velocity can guarantee the validity of
forward-scattering approximation. In order to keep this
approximation valid, we calculate phonon number $n_{ph}$ at $P=0$.
In ionic crystals, the phonon carried by electron is virtual when
the energy of moving electron is lower than that of optical phonon
which has a finite energy gap. Superficially for BEC system phonon
is gapless, no matter how small the velocity of the impurity is,
phonon is able to create. But it can also not be created for
arbitrary small impurity velocity due to above stated BEC mechanics.
So phonon number we calculate is also virtual. Corresponding
behavior for phonon number is showed in Fig.2 and is very similar to
the behavior of effective mass of impurity. This observation is
consistent with our intuition about polaron$\colon$ the more number
of the phonon encompassing the impurity, the larger the force
dragging the impurity and so the larger the impurity effective mass.

\begin{figure}[tbp]
\centering
\includegraphics [width=7.5cm, height=6.0cm]{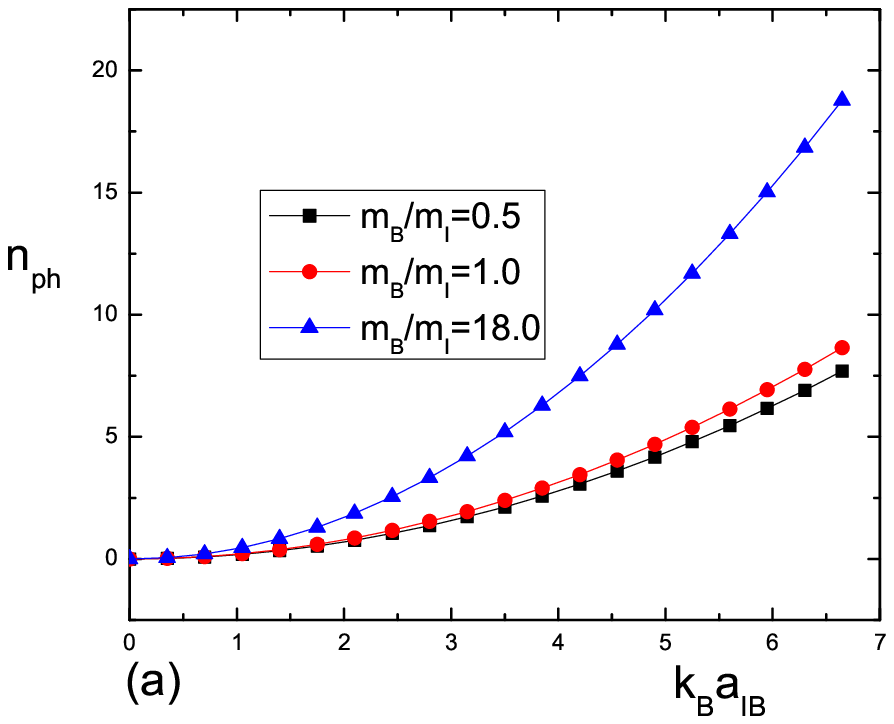}
\includegraphics [width=7.5cm, height=6.0cm]{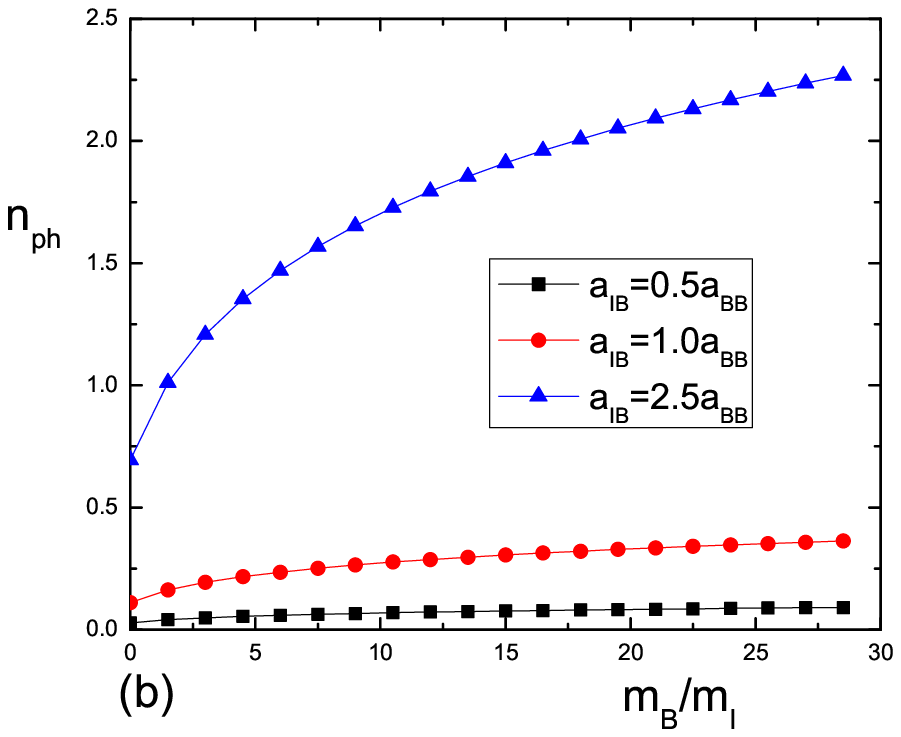}
\caption{The behavior of the phonon number $n_{ph}$ as a function of
boson-impurity scattering length $a_{IB}$ for different mass ratios
$m_{B}/m_{I}$ of boson to impurity (a); as a function of
boson-impurity mass ratio  $m_{B}/m_{I}$ for different
boson-impurity scattering length $a_{IB}$. We have chosen
$a_{BB}k_{B}=0.87$.} \label{fig.2}
\end{figure}

Below we calculate energy related to an impurity with zero velocity.
The whole energy of system is
\begin{eqnarray}
E_g(0) &=& E_B+E(0)+\mu n \nonumber \\
&\approx& (g n_0^2+E_B)+(\lambda n_0+E(0))\label{1.26}
\end{eqnarray}
Where we have used (\ref{1.4}) and make another approximation $\mu
n\approx\mu n_0$ which is valid for dilute BEC. The first term on
the right hand side in (\ref{1.26}) is the energy of pure bosonic
system, so the second term gives impurity energy in BEC system. The
calculation about energy must be careful. In Hamiltonian
(\ref{1.0}), we have used the contact interaction to represent the
true potential for which its Fourier transformation should fall off
at large momentum. This substitution leads to the ultraviolet
divergence of the ground state energy, so that a regularization must
be forced. In fact, this divergence is not fundamental and can be
regularized by way of absorbing more high order scattering
contribution into scattering length \cite{fetter}
\begin{eqnarray}
\frac{2\pi
a_{IB}\hbar^2}{m_{red}}=\lambda-\frac{2m_{red}\lambda^2}{\hbar^2}
\sum_k \frac{1}{k^2}   \label{1.27}
\end{eqnarray}
After taking these measures, the impurity energy $E_{I}$ after
scaled by $2\pi \hbar^{2} a_{BB} n/m_{B}$ is
\begin{eqnarray}
E_{I} = \rho(1+\frac{m_B}{m_{I}})\frac{k_B a_{IB}}{k_B
a_{BB}}+\frac{4\rho}{\sqrt{3}\pi^{3/2}}(1+\frac{m_B}{m_{I}})^2
\frac{(k_B a_{IB})^2}{(k_B a_{BB})^{1/2}}\Pi^{(2)}\label{1.28}
\end{eqnarray}
where
\begin{eqnarray}
\Pi^{(2)}= \int_{0}^{\infty} d y \,
y^2\left[\frac{1}{[1+(m_B/m_I)]y^2}-\frac{1}{(y^2+2\rho)^{1/2}
[(m_B/m_I)y+(y^2+2\rho)^{1/2}]}\right]\label{1.29}
\end{eqnarray}
The corresponding behavior of $E_{I}$ is plotted in Fig.3.

\begin{figure}[tbp]
\includegraphics [width=7.5cm, height=6.0cm]{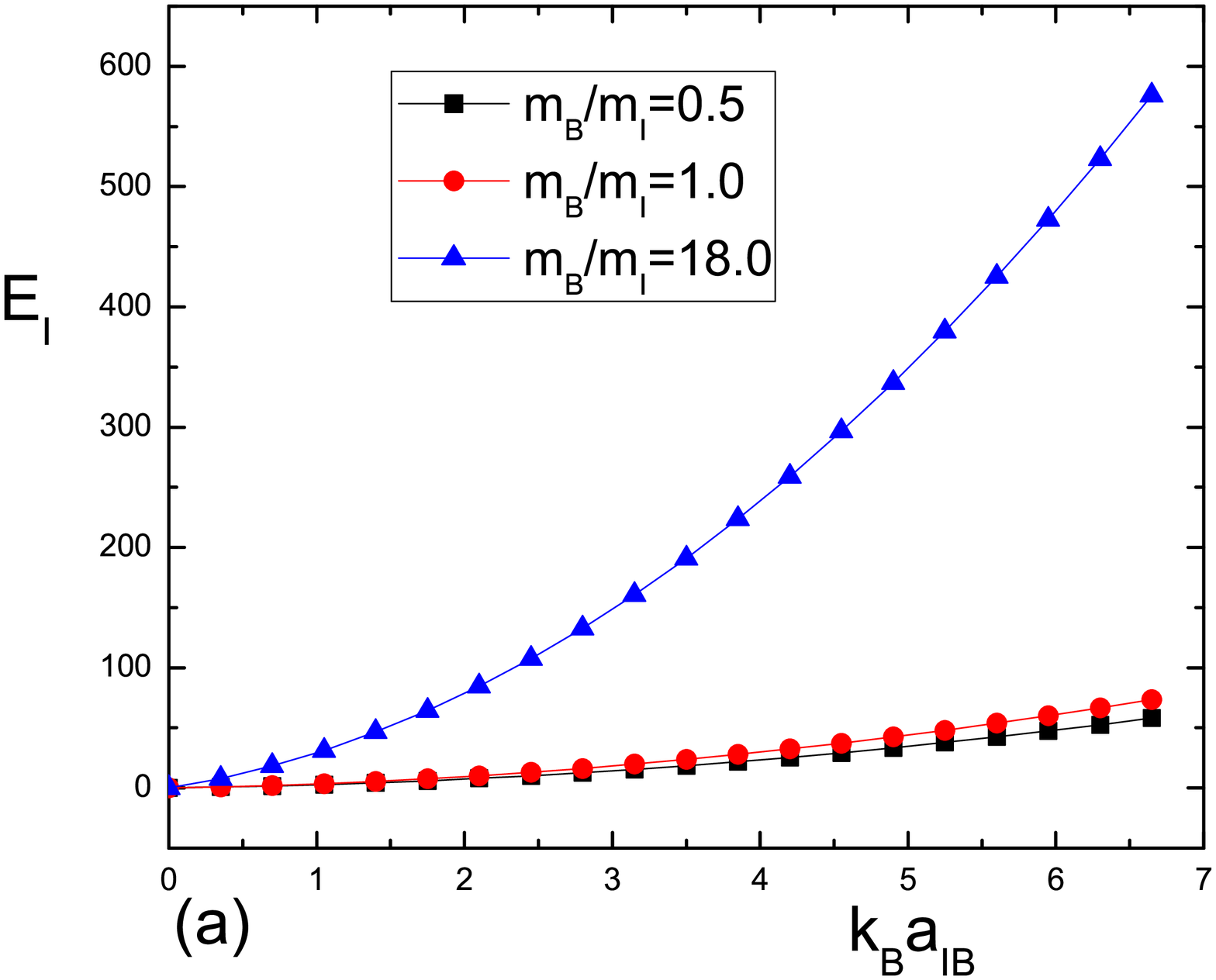}
\includegraphics [width=7.5cm, height=6.0cm]{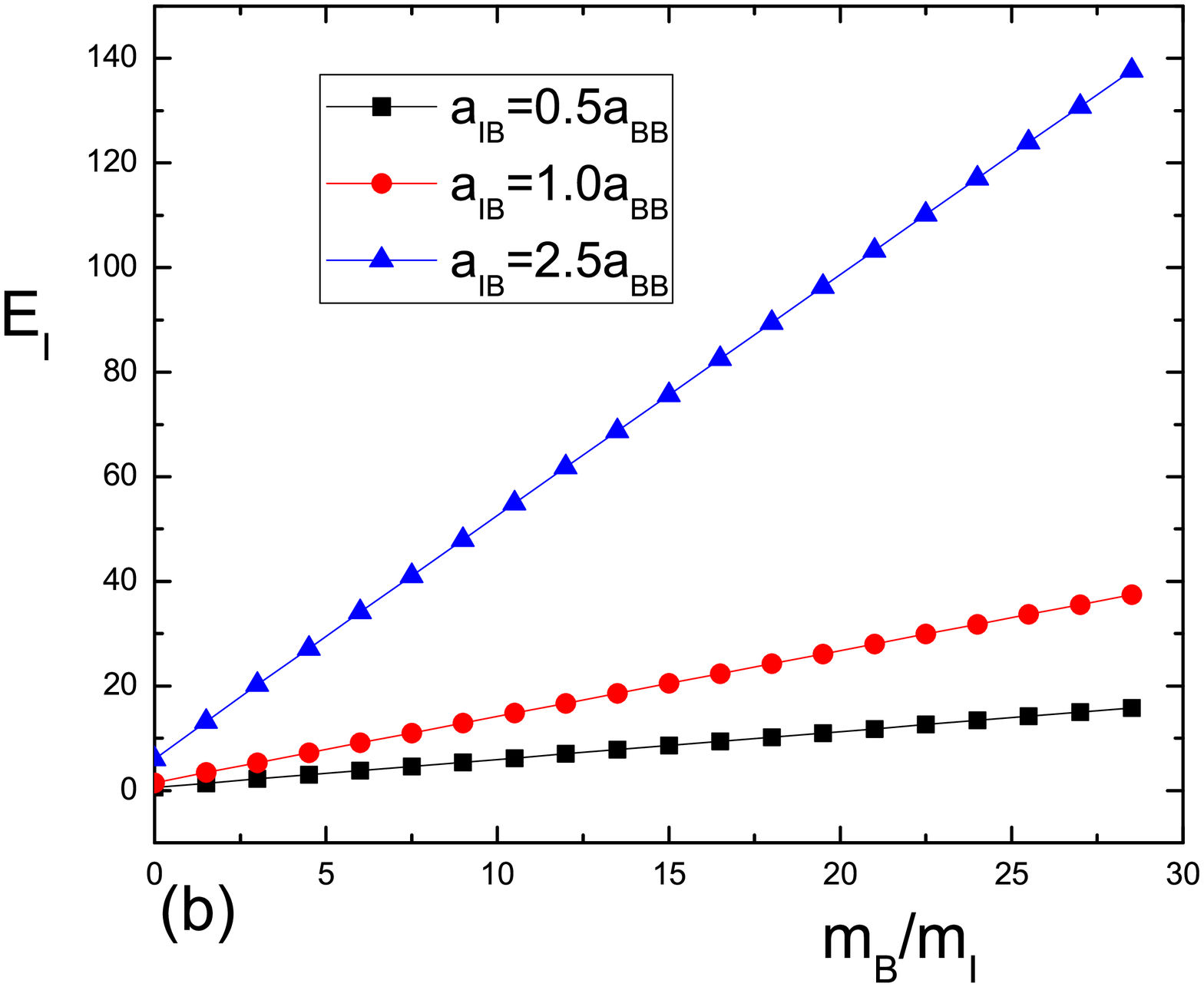}
\caption{The behavior of impurity energy $E_{I}$ as a function of
boson-impurity scattering length $a_{IB}$ for different mass ratio
$m_{B}/m_{I}$ of boson to impurity (a); as a function of
boson-impurity mass ratio  $m_{B}/m_{I}$ for different
boson-impurity scattering length $a_{IB}$. We have chosen
$a_{BB}k_{B}=0.87$.} \label{fig.3}
\end{figure}

\section{Discussion and Conclusion}

At last, we discuss the validity of our approximation (\ref{1.100})
in the frame of LLP theory. Following the method calculating
$n_{ph}$, (\ref{1.10}) equals to
\begin{eqnarray}
n = n_{0} + \sum_{k} v_{k}^{2} + \sum_{k} (u_{k}^{2} + v_{k}^{2})
\mid f(k) \mid ^{2} - 2 \sum_{k} u_{k} v_{k} f(k) f(-k) \label{1.29}
\end{eqnarray}
After some predigestion and dividing both sides of (\ref{1.29}) by
$n$
\begin{eqnarray}
1&=&\rho+\frac{8}{3\sqrt{6}\pi^{3/2}}\rho^{3/2}(a_{BB}k_B)^{3/2}+\frac{2\rho}{\sqrt{3}\pi^{3/2}n}
(1+\frac{m_{B}}{m_{I}})^{2} \frac{(k_Ba_{IB})^{2}}
{(k_Ba_{BB})^{1/2}} \Pi^{(3)}\nonumber
\\ &-&\frac{\rho^2}{2\sqrt{3}\pi^{3/2}n}\frac{m_{I}}{m_{B}}(1+\frac{m_{B}}{m_{I}})^{2}
\frac{(k_Ba_{IB})^{2}}{(k_Ba_{BB})^{1/2}}\frac{1}{\widetilde{P}(1-\eta)}\Pi^{(4)}
\label{1.30}
\end{eqnarray}
with
\begin{eqnarray}
\Pi^{(3)}&=&\int_{0}^{\infty}dy\frac{y^{2}+\rho}{y^{2}+2\rho}\frac{1}
{[(y^{2}+2\rho)^{1/2}+(m_{B}/m_{I})y]^{2}- 4(m_{B}/m_{I})^{2}
\widetilde{P}^{2}(1-\eta)^{2}} \nonumber \\
\Pi^{(4)}&=&\int_{0}^{\infty}dy\frac{1}{(y^{2}+2\rho)[\sqrt{y^2+2\rho}+\frac{m_B}{m_I}y]}
\ln \mid \frac{\frac{m_{B}}{m_{I}} y + \sqrt{y^{2} + 2
\rho}+2\frac{m_{B}}{m_{I}} \widetilde{P}(1 - \eta)}
{\frac{m_{B}}{m_{I}} y + \sqrt{y^{2} + 2\rho}-2\frac{m_{B}}{m_{I}}
\widetilde{P}(1 - \eta)} \mid \nonumber
\\ \label{1.31}
\end{eqnarray}
Equation (\ref{1.29}) is the most basic equation from which we can
study the effect of single impurity on condensate fraction. For
dilute bosonic system, the condensate fraction $\rho$ approaches
unit. The density of a typical BEC system in experiments is
$10^{13}-10^{15} cm^{-3}$. Although it is small in contrast to
typical density of gas, liquid and solid, it remains a large number.
From (\ref{1.30}) there is a factor $n$ in the denominators of last
two terms on the right hand side. If there are lots of impurities,
the last two terms would also include a factor proportional to
impurity density. So the last two terms are small and negligible in
contrast to the first and second terms in the light of the existence
of single impurity. In this way the approximation (\ref{1.100}) is
proven to be valid.

In conclusion, by making Bogoliubov and forward-scattering
approximations we have obtained an equivalent Hamiltonian which is
similar to the Fr\"{o}hlich large polaron Hamiltonian and calculate
effective mass of impurity, the phonon number carried by the
impurity and the impurity energy on a basis of LLP theory. In
addition, we also point out the valid range of forward-scattering
approximation that the velocity of the impurity must be smaller than
Landau critical velocity of BEC system.

\section*{Acknowledgement}

The work was supported by National Natural Science Foundation of
China under Grant No. 10675108.

\end{document}